\documentclass{aa520}
\usepackage{graphicx,natbib,epsfig}
\bibpunct{(}{)}{,}{a}{}{,}
\begin{document}
\title{The Sun as a MHD generator: application of a new heating mechanism for the coronal loops and closed
magnetic structures}
\author{D. Tsiklauri}
\institute{Joule Physics Laboratory, Institute for Materials Research,
School of Computing, Science and Engineering,
University of Salford, Salford, Greater Manchester, M5 4WT, United Kingdom.}
\date{Received xxxx / Accepted xxxx }

\abstract{We investigate a possibility of heating of the loops and other closed
magnetic structures in active regions of the solar corona 
by the flow of solar wind (plus other flows that may be present) across the magnetic field lines (that are 
perpendicular to the flow), 
in a similar manner as a conventional MHD generator works. 
A simple model is formulated which allows to calculate the typical 
currents generated by this mechanism. 
This enables to make a judgement whether the loops in solar active regions can be heated by
the proposed mechanism through the associated Ohmic dissipation. 
It is shown that even if the heating region width is of the order of 
a few proton Larmor radii (this effectively means that a loop is represented by
{\it nano-threads}, by analogy with the nano-flare coronal heating mechanism,
with width of a few proton Larmor radii) and the plasma flow speed is 50 km s$^{-1}$
then  only about 1\% of the heating requirement of coronal loops could be met
under this conjecture. 
Other implications of the model, such as a predicted spatial heat distribution
along the loop as well as application of the model to the other places e.g. chromosphere 
and corotating interaction region
structures in the solar wind are discussed.
 }
\titlerunning{The Sun as a MHD generator ...}
\authorrunning{Tsiklauri}
\maketitle

\section{Introduction}

The coronal heating problem, the puzzle of what maintains the solar corona 200 times
hotter than the photosphere, is one of the main outstanding questions in solar physics
(see e.g. \citet{a04} for a recent review on the subject).
\citet{a04} categorises theoretical models of coronal heating processes 
into 5 main groups, 
according to the main underlying or driving physical processes. 
These groups are (i) DC (Direct Current) stressing and reconnection models,
(ii) AC (Alternating Current) wave heating models,
(iii) Acoustic heating,
(iv) Chromospheric reconnection,
(v) Velocity filtration.
The main focus of the solar physics community research 
in the coronal heating theory has been concentrated on
the DC and AC type models. 

These two types 
describe the electro-mechanic coronal response to the photospheric 
driver that provides the ultimate energy source for heating. 
It is believed that the perturbations of 
magnetic field are generated e.g. by the convective motions in the solar
convective zone. Then the perturbations
propagate through the photosphere to the corona and ultimately constitute the main
energy source of coronal heating. These perturbations propagate with the Alfv\'en speed. 
The judgement whether the heating is of the DC or AC type can be made by 
comparing time-scales
of the heating with the Alfv\'en transit time along a coronal loop:
(i) When the photospheric driver, e.g. random motion of magnetic field line 
footpoints, changes the 
boundary condition on time scales much longer than the Alfv\'en transit time along a 
coronal loop, the loop can adjust to the changing boundary condition in a 
quasi-static 
manner. Hence the coronal currents are almost direct ones, 
which attributes a model to the DC type. 
(ii) However, when the photospheric driver changes faster than a coronal loop can 
adjust (e.g., by damping and dissipation of incoming Alfv\'en waves and 
fast and slow magnetosonic waves generated due their coupling to Alfv\'en waves
on plasma inhomogeneities), the coronal loop 
experiences an alternating current, which attributes a model to the AC type.

The DC and AC coronal heating models are subdivided further by possibilities of
how the currents are dissipated: 
either by magnetic reconnection \citep{su81,p83,p88,pht02,p03,plt03}, 
Ohmic dissipation via current cascading \citep{vb86,gn96}, and viscous 
turbulence \citep{hp92,evp96,ecc96} in the case of DC models, or 
by Alfv\'enic resonance, i.e. resonant
absorption \citep{i78,d87,pgk89,ods94,ods95,eg94,eg95,eg96,vh04,ru99}, 
phase mixing \citep{hp83,nph86,p91,nrm97,rnr98,rgr99,dmha00,bank00,tan01,hbw02,tn02,tna02,tnr03,tss05,tss05b}, 
and turbulence \citep{ip95,dmm01} in the case of AC models. 

As an alternative to current dissipation, some heating could also be produced by 
compressional waves (i.e., by acoustic waves or shocks \citep{kis81}).
It should be mentioned that a problem which models with 
compressional waves face is that such waves cannot propagate
from photosphere into corona due to the density gradient.
However, interest in these have been recently renewed due a
possibility of generation of fast and slow magnetosonic waves via their coupling to Alfv\'en waves
on plasma inhomogeneities \cite{nrm97,tan01,tn02,tna02,tnr03}.

In addition yet another approach is adopted by the chromospheric reconnection
models: the idea is that chromospheric reconnection through
generation of fragmented flux-tubes, strong currents, fast magnetosonic
waves, and upwards plasma flows can contribute to the coronal heating \cite{s99, rhw01,fs00,sky00,smk00,
smk01,sts01}.

Also a completely different physical 
mechanism has been proposed named velocity filtration, which is based on the influence of the 
gravitational potential field in the corona on a postulated non-Maxwellian chromospheric 
velocity distribution \citep{sc94}.   

In this paper we investigate yet another mechanism for the heating of solar coronal
loops and other closed magnetic structures in general. 
We study a possibility whether the loops could be
heated by the flow of solar wind plasma (plus other flows that may be present) across them by generating
currents in similar manner as a conventional Magnetohydrodynamic (MHD) generator works. 
In section 2 we formulate a simple model which allows to estimate typical 
currents generated by this mechanism. We show that 
if the heating region width is of the order of 
few proton Larmor radii and the plasma flow speed is 50 km s$^{-1}$ (which some observations do report) 
even then the heating of coronal loops could be only
1\% of the heating requirement under this mechanism. 
In section 3 we discuss some relevant aspects pertaining to the
studied mechanism.

\section{Main Consideration}

Similarly to the stressed-induced current cascade models (described e.g. in
\citet{a04}, p. 366), we try to estimate energy budget necessary to heat
the solar corona. It is assumed by the stressed-induced current cascade models
that the random footpoint motion stirs up a potential magnetic field, 
and hence non-potential fields and pertaining currents 
are generated. One can estimate a significance of the Ohmic (Joule) dissipation 
associated with these currents and whether they can 
contribute to the coronal heating in a following manner. 
The volumetric 
heating rate $E_H$ requirement for a loop with length $l$ which is heated 
at both footpoints in the units of erg cm$^{-3}$ s$^{-1}$ is 
\begin{equation}
E_{H0}=\frac{2 F_{H0}}{l}=2 \times 10^{-3} \left( \frac{l}{10^{10} \rm{cm}} \right)^{-1}
\end{equation}
Here we used $F_{H0} = 10^7$ cm$^{-2}$ s$^{-1}$ from 
Table 9.1 in \citet{a04}).
The required current density $J$ for the Ohmic dissipation, given by
$E_H = j^2/ \sigma$,  is hence,
\begin{equation}
J=\sqrt{\sigma E_{H0}}=1.1 \times 10^{7} 
\left( \frac{l}{10^{10} \rm{cm}} \right)^{-1/2} \;\;\;\;\; \rm{(esu)}.
\end{equation}
Here the classical conductivity of $\sigma = 6 \times 10^{16}$ s$^{-1}$ 
for a $T = 2$ MK corona has been assumed.

In general the geometry of a magnetic field in active regions is quite complex.
It mostly consists of closed structures such a loops, arcades, canopies, etc.
Instead of considering curved magnetic fields,
we simplify treatment by approximating active region loop 
with a straight slab of length $l$ ($l$ and associated with the loop
magnetic field are both along $z$-axis, $\vec B=B \vec z$)
and radial thickness $d$ ($d$ is along $x$-axis).
Note that $d$ plays the same role as a length (dimension along the plasma flow)
of a MHD generator \citep{kt73}.
The plasma flows along $x$-axis with velocity $\vec U=U \vec x$.
The sketch of the model is given in Fig.~1.
\begin{figure}[]
\resizebox{\hsize}{!}{\includegraphics{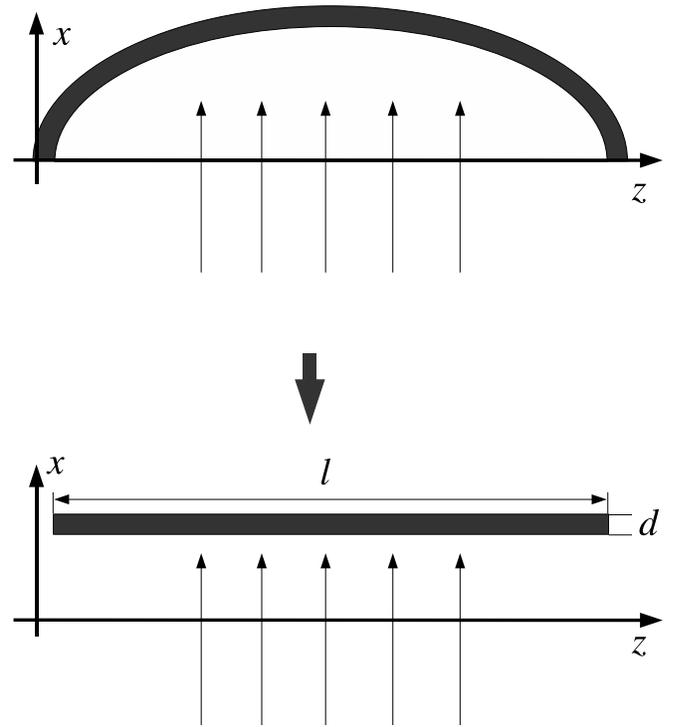}} 
\caption{The sketch of the model.}
\end{figure}
For our model we can write Ohm's law in the following form (e.g. \citet{kt73}):
\begin{equation}
\frac{\vec U \times \vec B}{c}=- \vec E+\frac{\vec J}{\sigma}+
\frac{|\omega_{ce}| \tau}{B \sigma} \vec J \times \vec B.
\end{equation}
Here we use standard notation with $\vec E$, $\vec J$ and $\sigma$ being
electric field, current density and plasma conductivity respectively.
$\omega_{ce}$ is an electron gyro-frequency and $\tau$ is time between collisions.
Note that we can omit Hall current, the last term, in the Eq.(3) 
because we assume
that circuit is open in the direction of plasma flow and no current can
be sustained. However, current due to the Lorentz force can flow
in $y$-direction (i.e. into the plane of Fig.~1.  Note: the generated
current flows not along the slab, but across it, in $y$-direction!
This is quite different from the stressed-induced current cascade models (described e.g. in
\citet{a04}, p. 366) in which {\it azimuthal} twisting of the magnetic field lines
of a cylindrical loop generates current {\it along} the loop. In our model current flows 
from one edge of the footpoint to the other edge of the same footpoint, but involving
entire bulk of the slab (which mimics the loop)).
In a steady state usual MHD equation of motion
\begin{equation}
\rho \frac{\partial \vec U}{\partial t}+\rho \, \vec U \cdot \nabla \vec U=
\frac{\vec J \times \vec B}{c},
\end{equation}
is reduced to the following form
\begin{equation}
\frac{\rho}{2} \frac{d U^2}{dx}=
\frac{|\vec J \times \vec B |}{c}.
\end{equation}
Note that we neglected pressure gradient in the Eq.(4) because dominant
force in the solar corona is associated with magnetic field, i.e. plasma $\beta \ll 1$.
We can simplify Eq.(5) by putting $d / dx \simeq 1 / d $
\begin{equation}
\frac{\rho U^2}{2}  \approx
\frac{|\vec J \times \vec B |}{c} d.
\end{equation}
Hence we can estimate the magnitude of the current density $J$ as
\begin{equation}
J=\frac{1}{2}\frac{\rho U^2 c}{B d}.
\end{equation}
Note that the same as Eq.(7) applies to the current density
produced by and MHD generator. However, for the latter $d$ is replaced
by the length of the MHD generator along which a plasma flows.

Based on Eq.(7) we can make following
estimates for the current density. We use 
following typical values (e.g. \citet{a04}) of parameters:
An electron number density of an active region loop
$n_e=2.0 \times 10^{9}$ cm$^{-3}$ (hence the density is $\rho=\mu m_p n_e$.
We put $\mu=0.61$ corresponding
to the a totally ionised gas composed of 90 \% of hydrogen and a 10 \% of helium \citep{sedp99});
In closed coronal magnetic structures, 
\citet{noci97}, using UVCS operating on board of SOHO, 
 showed that the spatial distribution of the r.m.s. velocities of OVI ions, in 
streamers, is characterised by the fact that in the brightest regions (the streamer core, around the
axis, at the lowest heights) the r.m.s. velocity is the smallest.
The r.m.s. velocity is roughly constant in the core while, outside it, it increases, both
moving in radial distance and in heliolatitude.
\citet{noci97}
measured r.m.s. speeds (in a core of
a helmet streamer) to be of the 
order of $\leq 50$ km s$^{-1}$ between  $1.5 -3 R_{\sun}$.
We use this value of $U=50$ km s$^{-1}$ in the Eq.(7), however this should be
used with caution that: what is actually measured is the r.m.s. speed i.e. some bulk motions of
plasma carrying OVI ions with it (through measurement of OVI $\lambda 1037/ \lambda 1032$ ratio)
which may not necessarily be the velocity of the
bulk plasma (protons and electrons) {\it across} the magnetic field of
the closed structures.
It should be also mentioned that some observations (see e.g. bottom right panel of Fig.~4 in \citep{stra02}) 
indicate that outflow speeds of O$^{5+}$ ions for the two UVCS slit orientations at 2.33 $R_{\sun}$ from the Sun's
centre i.e. 1.33 $R_{\sun}$ above its surface, above a helmet streamer, are less than $\approx 20$ km s$^{-1}$
in the central region of the streamer, which is in contradiction with \citet{noci97}'s measurements. 
Although \cite{of2000} argues that outflow speeds for the
bulk of the plasma (protons and electrons) should be higher than that for O$^{5+}$ ions (because the latter
drag out through Coulomb collisions), still plasma flows could be as low as $\approx 10$ km s$^{-1}$
(Ofman, private communication). Adopting such low values of the plasma outflow speeds 
would diminish our estimates considerably as the current is proportional to the $U^2$.
For our estimates, at any rate, we use more "optimistic" value of $U=50$ km s$^{-1}$ quoted above 
(also see our comment (ii) in the Discussion section).
A magnetic field of 100 Gauss. 
 Note that we only use longitudinal magnetic field strength, i.e. we totally ignore
azimuthal twist of the magnetic field.  There
is only evidence for weak twists in coronal loops,
of the order of $B_{\phi}/B_z \approx 0.01$, from measurements
as well as from the kink instability criterion. Thus, taking into account
only $B_z$ component of the magnetic field seems as a good approximation.
This is quite different from the stressed-induced current cascade models 
in which {\it azimuthal} twisting of the magnetic field lines
of a cylindrical loop generates current {\it along} the loop. In our case the current flows 
from one edge of the footpoint to the other edge of the same footpoint, involving
entire bulk of the slab (that mimics the coronal loop).
An additional thought is needed when specifying the parameter $d$.
Initially we can try putting an observed loop width.
\citet{ana00} present data based on 41 active region loops.
They found that typical loop width is few Mm.
Hence putting loop radial thickness $d \approx 1$ Mm we obtain.
\begin{equation}
J=0.08 \left( \frac{d}{1 \rm{Mm}} \right)^{-1} \;\;\;\;\; \rm{(esu)}.
\end{equation}
This estimate turns out to be about eight orders of magnitude
less than the requirement to heat solar corona cf. Eq.(2).

In the next estimate, following \citet{ana00}, we adopt the
multi-tread model of a coronal loop. A good graphical representation of this idea
can be found in Fig.~1 from \citet{ana00}, in which a loop is represented by
100 fine loop threads with random footpoint positions and random fluxes. 
Thus assuming value for $d=0.01$ Mm, our estimate of current density
will still fall six orders of magnitude less than the coronal heating requirement.

It has been a general trend that as the resolution of EUV telescopes
increased more finer and finer structures were observed.
It is a general agreement now that loops have much finer substructure than
previously thought. The lowest possible scale for the MHD approximation 
to be valid is few Larmor radii of protons.
Which is $r_L=V_\perp / \omega_{cp}$.
$\omega_{cp}= e B / m_p c$ is a proton gyro-frequency.
Putting $V_\perp=\sqrt{3 k_B T/m_p}$ with temperature of 
$T = 2$ MK and $B=100$ Gauss, $r_L$ is then $23.23$ cm (\citet{st01} give similar
estimate for $r_L$ within a factor of $\sqrt{3}$).
Therefore assuming that loop consists of substructures, what we
call { nano-threads (by analogy with the nano-flare coronal heating mechanism)}), with a width of a few proton Larmor radii 
for the generated current we obtain
\begin{equation}
J=1.1 \times 10^5 \left( \frac{d}{3 r_L} \right)^{-1} \;\;\;\;\; \rm{(esu)}.
\end{equation}
This is only about 1\%  what is required for the heating of solar corona cf. Eq.(2), which
obviously cannot be regarded as a sizable contribution to the heating.
By obtaining this, one should be also aware of other considerations:

(i) One may question a possibility of existence of
such nano-treads. 
 There is no direct evidence in the observations
that the coronal loops consist of threads with
width of a few Larmor radii. In fact, essentially all
finest loops observed with TRACE have a width
of order 1000-2000 km and have a monolithic
structure, that is a single temperature at any
cross-section, which {\it presently} does not support the
nano-thread concept. 
The TRACE CCD camera has 0.5 arcsec pixels which is 366 km on the sun.
Hence it is possible that the smallest observed 1000 km wide
monolithic structures are simply too close to the resolution limit
and future high spatial resolution
space missions may or may not 
reveal further sub-structuring ultimately
corroborating or disproving our nano-thread conjecture.

Situation here is possibly similar to the
stressed-induced current cascade model, which based on the
random footpoint motion (by twisting the field lines) predicts a 
current density, 
which is about 5 orders of magnitude smaller than required to satisfy 
the coronal heating requirement Eq.(2). \citet{a04} concludes then that 
the Joule dissipation is generally inefficient in the corona, unless $10^5$ 
smaller transverse length scales can be produced. This was the main 
motivation of the model by \citet{vb86}, who proposed a 
current cascade model, where free magnetic energy is transferred 
from large to small length scales in the corona as a result of 
the random motion of photospheric footpoints.

(ii) The other mechanisms reviewed in the introduction section
are main contributors to the heating budget. And MHD generator mechanism
contributes only tiny 1\% even at the smallest possible scale.

\section{Discussion}

The main novelty of this study is exploration of a conjecture
that loops in the solar corona could be heated by
currents which are generated by solar wind (plus other flows that may be present)
across the magnetic field lines in similar fashion
as conventional MHD generator works.
In previous sections we gave estimates of generated currents and
appropriate heating scales.

We would like to close the paper by discussion of
following four points.

(i) Spatial distribution of heat deposition along the
loop holds the key to an understanding of the physical precesses
that are responsible for heating of solar corona in general.
This quantity can be inferred e.g. from high resolution images
produced by the Yohkoh Soft X-Ray Telescope, the SOHO 
Extreme-Ultraviolet Imaging Telescope and the Transition Region
and Coronal Explorer (TRACE). \citet{ana00}
established that long loops $> 100$ Mm are
preferentially heated near footpoints
with the heating scale height of about 20 Mm and are far from
hydrostatic equilibrium.
While shorter loops $\simeq 20$ Mm are heated almost
uniformly and are predominantly in hydrostatic equilibrium.
Although the heating budget requirement is not
met by our model, it predicts that loops with weak curvature
(radius of curvature is much greater than loop length, $R_{\rm curv} \gg l$)
will be heated almost uniformly, as the most part of
the loop (with its magnetic field) will be at $90^\circ$ angle
to the plasma flow. Thus, $\vec U \times \vec B$ term will be
significant in all parts of the loop.
This is actually what is seen in the \citet{ana00}
data: e.g. if we look at images of loops 2,3,4 (length $\simeq 20$ Mm)
of the TRACE 171+195 data they all have large curvature radii (are almost flat)
and are heated uniformly, i.e. in accord with our model prediction.
However, as we estimated above heating requirement is still not met by our model. Hence,
the discussion about the spatial distribution of heat deposition along the
loop has only academic interest.
For the loops with strong curvature
(radius of curvature is of the order of loop length, $R_{\rm curv} \simeq l$)
our model predicts that footpoint parts will not be heated as the
magnetic field is almost parallel to the plasma flow and thus
at footpoints $\vec U \times \vec B$ term will be
small and only the apex part of the loop will be heated.
If we look at images of loops 30,31,33 (length $\simeq 200$ Mm)
of the TRACE 171+195 data they all have curvature radii comparable to the
loop length (are quite curved)
and are heated near footpoints. This faces an additional problem to our model
which predicts no heating at footpoints and heating concentrated at the
apex for the reasons explained above. Thus, loops with strong curvature
are probably heated by some other mechanism (near the footpoints).
However, as for the coronal heating problem itself this does not
pose a serious problem as the number of 
short ($\simeq 20$ Mm) and flat loops is more significant as compared
the long ($> 200$ Mm) and curved loops, which are more rare in occurrence.

(ii) We also would like to comment on the value of density used in our
estimates of the magnitudes of currents generated by our mechanism.
We used the observational value of 
$n_e=2.0 \times 10^{9}$ cm$^{-3}$. 
In a conventional MHD generator value of density is prescribed by
the strength of the magnetic field across the flow (which tends to resist to the flow)
and the strength of the gradient of the plasma flow through the generator.
In the case of coronal loops however there are also up-flows of plasma
from the chromosphere and transition region. 
Also, \citet{sfhh98} suggested that in active regions
there are loop systems that are 
embedded in hot coronal structures looking in soft X-rays like fans of coronal rays. 
These structures are formed during the flare and extend high into the corona. 
\citet{sfhh98} found that such 
structures create mass flow from the active region into 
interplanetary space
and these are the sources of a density enhancement in the solar wind. 
This suggestion is supported by synoptic maps of solar wind 
sources constructed from scintillation measurements which show a source of enhanced 
solar wind density at the position of the active region.
It is not entirely clear
at this stage what effect the above complex factors 
could have on our simple calculation, but
we believe in the overall correctness of our model and its 
predictions. At any rate one can conjecture that the observed 
density in the active regions
of $n_e=2.0 \times 10^{9}$ cm$^{-3}$ is caused by the fact that
the closed magnetic field lines, transverse
to the plasma flow, resist the flow making it dense, such
that the observed density is obtained. The same occurs
in an MHD generator.

(iii) We would like to point out the difference between
our model and the AC and DC models mentioned in the introduction.
For the latter two the main source of energy is the convective
motions that drive either reconnection events or generate
waves which then dissipate. In our case, however, the source of
energy is in the plasma flow across the magnetic field lines,
i.e. solar wind plus other flows that may be present.
Commenting on possibility of actual flow across the magnetic field
in the coronal conditions we would like to stress that the parameter
which controls the possibility is $\sigma$, the plasma conductivity.
Since we do not use any enhanced values for $\sigma$ and the 
bulk flows in the cores of helmet streamers are observed \citep{noci97},
we believe that the flows across the magnetic field are possible.
Note that smallness of plasma-$\beta$ in the corona is not a problem
for our model
as this parameter controls only {\it compressibility} of plasma, not
the cross-field flow control.

(iv) Also, we would like to point out a possibility to apply our
mechanism in the
chromospheric footpoints of the loops or in the corotating interaction region
structures in the solar wind.
For example, in the chromosphere the neutrals are abundant and
the frozen-in condition, which inhibits the cross-field plasma flow can perhaps 
be violated. 
Possible across-the-field flows are the Evershed flows in
sunspots. Also, various granulation motions can perhaps be transverse to
the field and hence produce the current.
Although, in the chromospheric footpoints
there are only very small random motions detected
in horizontal direction (and thus perpendicular
to the vertical magnetic field), of the order of
 10 km s$^{-1}$, which yields currents four orders of magnitude below
the coronal heating requirement. But, on positive side
density in the chromosphere is much higher, hence
the value of the generated current could be enhanced by orders of magnitude.
Also, further out in the corotating interaction region
structures \citep{gos96}, in the solar wind, the magnetic field has a transverse 
to the plasma flow component and hence currents can be generated.

We also would like to estimate how "efficient" is the Sun
as a MHD generator: The dimensionless ratio which quantifies this, is
$N=\sigma B^2 d / (\rho U c^2)$ (in CGS units). For the most "optimistic" estimate
(Eq.(9)) along with the classical conductivity of $\sigma = 6 \times 10^{16}$ s$^{-1}$ 
for a $T = 2$ MK corona, $N=4.6 \times 10^9 \gg 1$. This means that, on one hand, the Sun
appears to be quite efficient MHD generator, as for typical ionised gas MHD generators
$N\approx 10$ \citep{hey68}. On the other hand, however, despite of this, coronal heating budget is still 
not met by our mechanism.

Interestingly, our conjecture with MHD generator can perhaps also lead to yet another
interesting possibility: if instead of generating current across  the loop (not along the loop!) via the flow of plasma
across the magnetic field, an electric field (created e.g. by some reconnection type event) 
can be applied across $y$-coordinate
into the plane of Fig.~1 i.e. from one edge of the footpoint 
to the other edge of the same footpoint, involving
entire bulk of the slab (that mimics the coronal loop), 
then the Lorentz force reverses its direction (see Fig.~3 in \cite{hey68})
and the system can act as a {\it MHD pump}, i.e. accelerate the plasma flow.
Whether this yet another conjecture is of any relevance to the solar wind
acceleration remains yet to be seen.

\begin{acknowledgements}
The author kindly acknowledges support from Nuffield Foundation 
through an award to newly appointed lecturers in Science,
Engineering and Mathematics (NUF-NAL 04).
The author would like to thank L. Ofman and V. Nakariakov for critical comments; L. Strachan for the 
additional data from his paper; the Editor, W. Schmidt, for useful comments, 
and B. Roberts for the encouragement and good advice.
\end{acknowledgements}
\bibliography{ms3167}
\end{document}